\documentclass[]{spie}  %>>> use for US letter paper
\pdfoutput=1
\usepackage[]{graphicx}
\usepackage{times}

\title{Data Management at the UKIRT and JCMT}

\author{Tim Jenness\supit{a} and Frossie Economou\supit{a}
\skiplinehalf
\supit{a}Joint Astronomy Centre, 660 N.\ A`oh\={o}k\={u} Place, Hilo, HI, 96720, U.S.A.\
}

\authorinfo{Further author information: (Send correspondence to T.J.)\\T.J.: E-mail: t.jenness@jach.hawaii.edu}
\begin{document}
  \maketitle

%%%%%%%%%%%%%%%%%%%%%%%%%%%%%%%%%%%%%%%%%%%%%%%%%%%%%%%%%%%%%
\begin{abstract}
  For more than a decade the Joint Astronomy Centre has been
  developing software tools to simplify observing and make it possible
  to use the telescopes in many different operational modes. In order
  to support remote operations the data handling systems need to be in
  place to allow observation preparation, flexible queue scheduling,
  data quality pipelines and science archives all to be connected in a
  data-driven environment. We discuss the history of these
  developments at UKIRT and JCMT and how the decision to combine
  software development at both telescopes led each to get features
  that they could not have justified if they were treated
  independently.
\end{abstract}

%>>>> Include a list of keywords after the abstract


%%%%%%%%%%%%%%%%%%%%%%%%%%%%%%%%%%%%%%%%%%%%%%%%%%%%%%%%%%%%%
\section{INTRODUCTION}
\label{sec:intro}  % \label{} allows reference to this section

The Joint Astronomy Centre runs two distinctly different
telescopes. The James Clerk Maxwell Telescope (JCMT) is the world's largest
sub-millimetre telescope with a 15-m primary mirror. The United
Kingdom Infrared Telescope (UKIRT) is a 3.8-m infrared telescope. Both are
sited on Mauna Kea in Hawaii and both telescopes have been operating for
many years (UKIRT since 1979 and JCMT since 1987) and developed many
legacy systems. In the late 1990s the software groups were merged and
code re-use and shared support became an important driver for the
future.

\section{DATA PIPELINES}

In the mid-1990s data reduction software was written directly for the
instrument with no intent to re-use when the instrument was
retired. At UKIRT CGS4DR\cite{1992ASPC...25..479S,1995ASPC...77..375D}
delivered near-publication quality data for
CGS4\cite{1990SPIE.1235...25M,1993SPIE.1946..547W} and was well
regarded at the time. Unfortunately its design was inflexible to
change and was highly dependent on the format and headers associated
with CGS4 data. At JCMT SCUBA\cite{1999MNRAS.303..659H} was delivered
in 1996 with a real-time data reduction system running on a VAX/VMS
system that was highly integrated with the observing system but had no
offline capability at all. Adding new features such as sky subtraction
became difficult to debug and the ability to regenerate the map
offline became very important as new CPU-intensive map-making
algorithms were developed. The resulting package,
SURF\cite{1998ASPC..145..216J}, could be used by the observer at the
summit to try to do a better job at making the map but shell scripts
had to be written to bulk process observations. The JCMT heterodyne
systems\cite{1990NIMPA.293..148B,1998SPIE.3351..148F} were a little
better off because the correlators (for example AOSC and the DAS) provided a
buffer between the users and the frontend but the extent of the data
processing was limited and a specialist offline tool (SPECX) reading a
niche file format (GSDD
\cite{1986QJRAS..27..675.,sun229})\footnote{The General Single Dish
  Data format (GSDD) was an offshoot of a joint MRAO, NRAO, IRAM
  project to provide a generic data format for single-dish
  telescopes.} was provided to allow the observer to process their own
data in near real-time if they were so inclined.

During this time software development effort in the UK moved from
VAX/VMS to Solaris and Digital Unix and budgets became tighter so
there was an opportunity to re-implement software systems whilst
reducing the cost for delivering data reduction software with new
instruments. The
ORAC\cite{1998SPIE.3349..184B,2000SPIE.4009..227B,2001ASPC..238..137W}
project at UKIRT was designed to rewrite the observatory
control system to provide instrument-agnostic interfaces for
observation preparation, sequencing and data reduction. The data
reduction part of the project was labelled
ORAC-DR\cite{1999ASPC..172...11E,2008AN....329..295C} and the aim was
to provide an infrastructure for data processing using reusable
processing blocks, called primitives, that represent high-level
concepts such as ``flatfield these images'' or ``combine into a single
image'', and are combined into ``recipes''. The system is designed to be
completely generic and to ensure this the initial development included
support for CGS4 and UFTI\cite{2003SPIE.4841..901R} at UKIRT and SCUBA
at JCMT. The overall design is shown in Fig.\ \ref{fig:train} 1 and
consists of the following features:

\begin{figure}
\begin{center}
\includegraphics[width=0.8\textwidth]{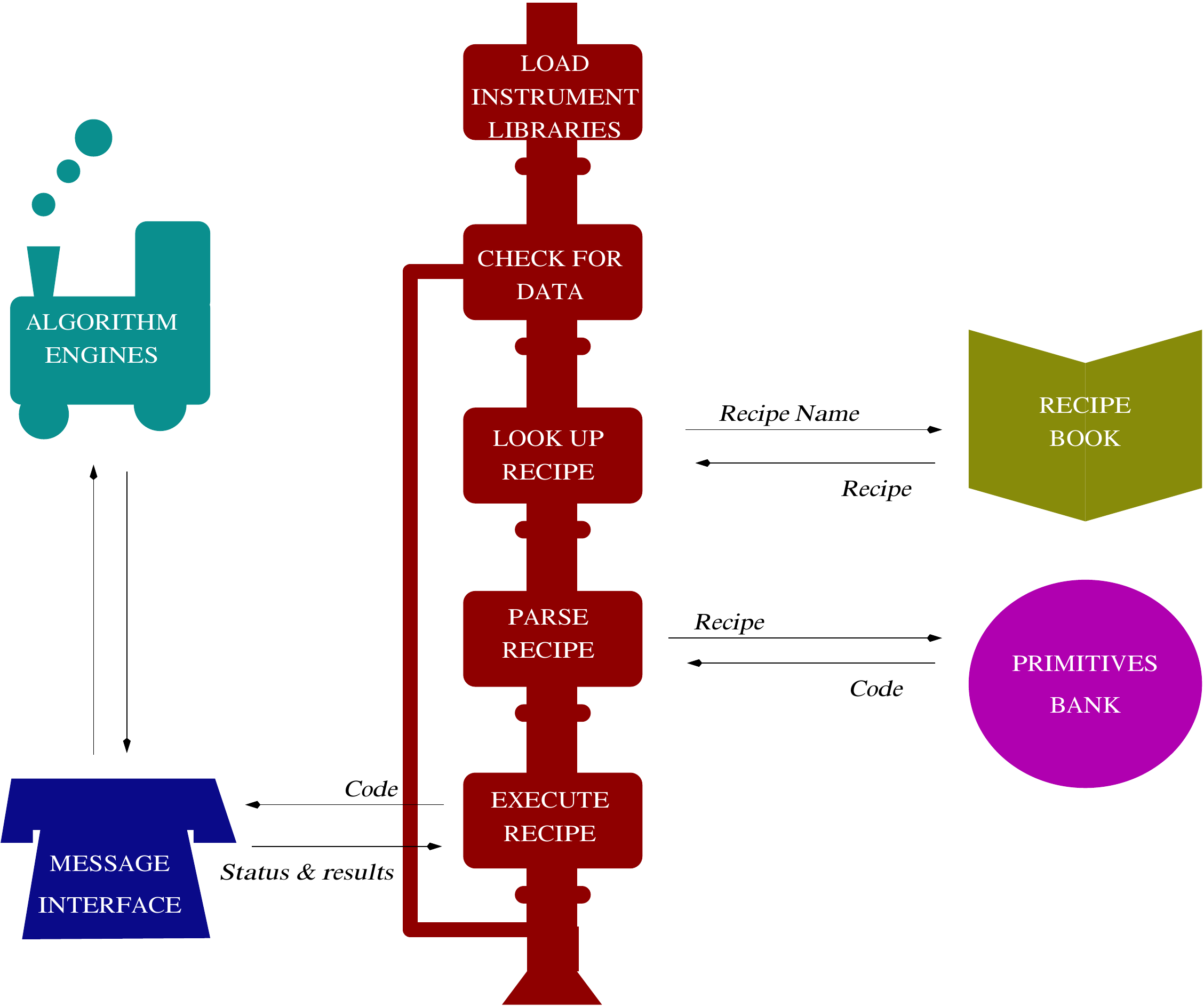}
\caption{\label{fig:train}
The design of the ORAC-DR pipeline.
}
\end{center}
\end{figure}

\begin{itemize}

\item Recipes are designed to be human readable and amenable to being
  built up by recipe builder software. No code should appear at the
  recipe level although primitives are allowed to take parameters to
  control behaviour. Additionally a recipe parameter file can be
  supplied to the pipeline to control global behaviour without
  requiring the recipe to be edited.

\item Primitives are written in code (in this case Perl) and should
  implement a simple well-understood concept.

\item The parser builds up an executable recipe from the
  primitives. Primitives are read at run-time and cached. If a
  primitive changes the recipe will change without having to rerun the
  pipeline. Additionally the parser automatically inserts status
  handling, debugging and timing code.

\item Low-level pixel processing is implemented by applications in the
  Starlink software
  collection\cite{1986BICDS..31...69D,1993ASPC...52..229W,2000ASSL..250...93W,2009ASPC..411..418J}
  controlled using the ADAM message
  bus\cite{1990NIMPA.293..148B,1992ASPC...25..126A}. Starlink was
  chosen because of the robust messaging system and well-defined error
  handling as well as support for variance propagation and automatic
  history writing\cite{sun33}. At the time, 1997,
  IRAF\cite{1993ASPC...52..173T} was not as robust and native support
  for numeric arrays in languages such as Perl or Python was very
  new. The pipeline architecture allows any messaging system and
  algorithm engine to be used and DRAMA\cite{1995SPIE.2479...62B}
  support has been implemented as an alternative to ADAM;
  demonstrating flexibility.

\item Pipeline state is stored in ``frame'' objects associated with
  each observation. This class changes depending on the instrument and
  handles file naming conventions, knowledge of file structures and
  metadata handling. There are also classes for handling a collection
  of frames, a ``group'', and calibration systems. The calibration
  classes, for example, know which flatfield or dark is relevant for a
  particular frame or how extinction correction should be applied.

\item As far as possible the primitives using generic metadata rather
  than using the actual FITS header stored in a particular file. When
  a file header is read the header is translated\footnote{See
    the \texttt{Astro::FITS::HdrTrans} Perl module at
    \texttt{https://github.com/bradcavanagh/perl-Astro-FITS-HdrTrans}}
  and both the original and translated header are stored in the frame
  object.

\end{itemize}

The most important design consideration is that the pipeline should
never pause to wait for input from the user. All the metadata needed
to process the observation must be available in the data headers and
they must be correct.  Algorithms must also be developed to do do as
good a job as possible without needing every recipe to be tuned in
advance. For example, the ACSIS\cite{2009MNRAS.399.1026B} pipeline
does not need to be told where the baseline is when doing baseline
subtraction. All that is required is that the pipeline is given an
idea of whether it is processing broad line, narrow line or line
forest data and it picks an appropriate algorithm. The narrow line
recipe uses the CUPID\cite{2007ASPC..376..425B} clump-finding software
to first find the emission regions before generating a mask that is
used to do high quality baseline subtraction.

In the 13 years since the pipeline was released it has been used for
CGS4, IRCAM, UFTI, UIST\cite{2004SPIE.5492.1160R},
MICHELLE\cite{1997SPIE.2871.1197G} and WFCAM\cite{2007A&A...467..777C}
on
UKIRT\cite{1999ASPC..172..175C,2001ASPC..238..314E,2003ASPC..295..237C};
SCUBA, ACSIS and SCUBA-2 on
JCMT\cite{1999ASPC..172..171J,2000ASPC..217..205J,2005ASPC..347..585G,2008ASPC..394..565J};
and IRIS2\cite{2004SPIE.5492..998T} on the
AAT\cite{2003ASPC..295..237C}. Proof-of-concept recipes have also been
developed for ESO and Gemini
instruments\cite{2004ASPC..314..460C,2005ASPC..347..580C,2003ASPC..295..237C}.

\section{FLEXIBLE SCHEDULING}

Submillimetre observing is very sensitive to the weather conditions
and in particular the amount of water in the atmosphere. In the first
few years of JCMT observing a ``classical'' scheduling approach was used
where a few shifts were allocated to an observer and they would get
their data or not depending on the weather. In some cases they would
observe low frequency projects in the best conditions or even observe
their own backup project if the weather was poor. It was realised that
this approach was not the most efficient use of the telescope and
plans were adopted to start a flexible approach to observing where the
people wanting the best conditions would be asked to come out to
observe but they would be given more nights than they needed and be
allowed to choose when to take their own data if they agreed to take
data for someone else if the weather was not good
enough\cite{1996ASPC...87....3R}. In 1996 JCMT began a flexible
observing system\cite{1996ASPC...87..168P} that involved multiple
projects being allocated blocks of time allowing the flexing inside
those blocks. As more experience was gained an implementation of this
scheme required the PIs to fill in a form saying what they would like
to observe and the observer going through a folder looking for a
relevant backup proposal.\cite{1998SPIE.3349..126W} Of course this
system had a large overhead for everyone and sometimes led to heated
email exchanges as the remote PI complained that their data was not
taken with as much care as would have been the case if they were
present. This was followed by a prototype system at JCMT that stored
all the targets to be observed in a database along with weather
requirements and allocated priority. No attempt was made to be able to
send the results of a query to the telescope so it was still necessary
to define the observation at the telescope. Meanwhile, UKIRT had been
thinking about queue scheduling for some
time\cite{1996ASPC...87..162B} and were running a service programme
based on a PostgresQL database that could track whether the data had
been accepted by the PI.

In June 1998 an initial project plan for a JCMT Observation Management
Project\cite{2000ASPC..216..101T} was developed but the full
implementation stalled as development effort was transferred to other
higher priority projects. By late 2000 it was clear that a single
unified system should be adopted by both telescopes to minimise the
overhead of supporting two disparate systems (the JCMT and UKIRT
software groups had recently merged) and to provide a system that
could not only tell you what to observe but also send the observation
to the telescope without the observer having to do anything. The JAC
Observation Management Project (OMP)\cite{2002ASPC..281..488E} was the
result of this merger heavily influenced by the earlier JCMT OMP and
the developments at UKIRT delivered as part of the ORAC project. The
requirements for the OMP were:

\begin{itemize}

\item An observation database populated by the astronomers using an
  observation preparation tool.\cite{2002ASPC..281..453F} The science
  programme is split up into scheduling entities known as Minimum
  Schedulable Blocks (MSBs).

\item A tool enabling the database to be queried and the results to be
  constrained by the current weather conditions and sorted by
  allocation priority.\cite{2002ASPC..281..144R}

\item A translation layer that will convert the abstract science
  description of the MSB into something understood by the telescope
  observatory control system (OCS).

\item A queue that will send translated output to the OCS one
  observation at a time and realise when an MSB has been
  completed.\cite{2004SPIE.5496..718K} The queue then gives an
  opportunity for the observer to accept or reject the MSB and
  associate this with any comment on data quality.

\item A collection of dynamically generated web pages (the ``feedback
  system'') to allow the remote observer to check on program
  completion, to comment on data quality and to retrieve the data
  files.\cite{2004ASPC..314..728D} This system also lets the support
  astronomer contact the PI to provide feedback and suggestions on the
  observing plan.

\item Data reduced by the ORAC-DR pipeline should be made available to
  the PI over the web as thumbnail images of maps or spectra, building
  on the WORF system deployed for the older
  systems\cite{1996ASPC..101..384E,1997ASPC..125..401J}.

\item Basic scheduling logic to allow MSBs to be disabled if other
  MSBs were observed first. This was implemented using \texttt{OR} and
  \texttt{AND} folders. An \texttt{OR} folder could be configured
  containing a mix of MSBs and \texttt{AND} folders containing
  MSBs. If \textit{N} of them were observed the remaining entries
  would be disabled.

\item Importantly for UKIRT the system had to work in purely classical
  mode if required.

\end{itemize}

The UKIRT ORAC system had been delivered with an observation
preparation tool (OT)\cite{2001ASPC..238..137W} derived from an early
release of a (since completely replaced) Gemini
OT\cite{1997SPIE.3112..246W} so it was decided early on in the project
that JCMT should adopt the same system and extend the XML schema with
features to support site quality (submm opacity, seeing, moon state),
logic folders and MSB folders.

UKIRT tested the full system in March 2002 to do their service
programme and the full system was live on JCMT by July 2002 for
SCUBA. The project was completed in September 2003. The heterodyne
instruments took a bit longer and an interim system was developed to
allow the OMP to be used with the legacy interfaces. The translation
system for the DAS backend would pop up a list of instructions for the
telescope operator to follow (e.g. tune to this frequency, set up a 5
x 5 grid, use the 250 MHz correlator settings) and when they were
finished they would click a button to tell the system that the MSB had
been observed. This was in use until ACSIS was commissioned in 2005/6.

The OMP system changed the way both telescopes ran and resulted in
improved completion statistics of highly ranked
projects.\cite{2002SPIE.4844...86R,2004SPIE.5493...24A}

\section{{eSTAR} RAPID RESPONSE}

Once we had a data reduction pipeline and a flexible scheduling system
it became possible for the telescope to be used in novel ways not
previously considered. One of our PIs was interested in fast followup
of gamma ray bursts (GRB) and had been talking to the eSTAR
project. eSTAR\cite{2004SPIE.5496..313A} is a system of autonomous
software agents designed for use in robotic telescope networks. The
system worked by having one agent monitoring GRB events, for example
from SWIFT, and it would then send out an observing request to
telescope agents asking whether they could do an immediate followup
observation. Any number can respond with a ``yes'' and the agent would
then send the request and wait to be told that the data was available
for analysis. The ultimate aim was that the agent would analyse the
data from the telescopes and then decide whether further observing was
warranted.

By early 2003 the eSTAR system was working for test robotic telescopes
but it was realised that a UKIRT telescope running the OMP and ORAC-DR
could be made to look identical to a robotic
telescope\cite{2006AN....327..788E}. The system worked by installing
an ``embedded agent'' at the JAC to handle the eSTAR requests. The
PI prepares a science program that has a dummy target position but
fully describes the rest of the observing parameters. When the agent
receives a request from a particular eSTAR user agent the user
information is translated to a local OMP project code and the agent
can check to see if time remains on the project for observing and if
the source is going to be up in a few hours. If the user agent
indicates that the embedded agent should do the observation the
template science program is retrieved, the MSB is copied and the
relevant target is inserted. UKIRT is not robotic and may well be in
the middle of an MSB and so can not instantly slew to the GRB
target. What actually happens is that an alert goes off in the control
room indicating a time-sensitive observation has entered the observing
database and the operator can decide whether to abort the current MSB
or continue to observe it.

Once the decision to observe it is taken the operator can locate a
relevant guide star and adjust exposure times as necessary. The final
piece of the puzzle is that the observing system writes a special
header item indicating that the files are associated with an eSTAR
trigger and storing the unique eSTAR identifier for the
observation. When the data reduction pipeline reduces the data at the
summit and generates the mosaic it sends the reduced data to the
embedded agent which forwards the result back to the PIs user agent
which can then decide whether to schedule a followup.

The first demonstration of eSTAR at UKIRT was carried out in August
2003 and the system triggered an automatic follow up observation of
GRB\,050716 in 2005\cite{2005GCN..3632....1T}. The observation was
queued in the observing system 48 seconds after it was detected by
SWIFT but due to operational difficulties the telescope could only
observe 56 minutes after initial detection. In 2009 GRB\,090423 was
detected within 20 minutes of the alert using
eSTAR\cite{2009GCN..9202....1T} (the weather was not good so it took a
while to open the telescope).

\begin{figure}
\begin{center}
\includegraphics[width=\textwidth]{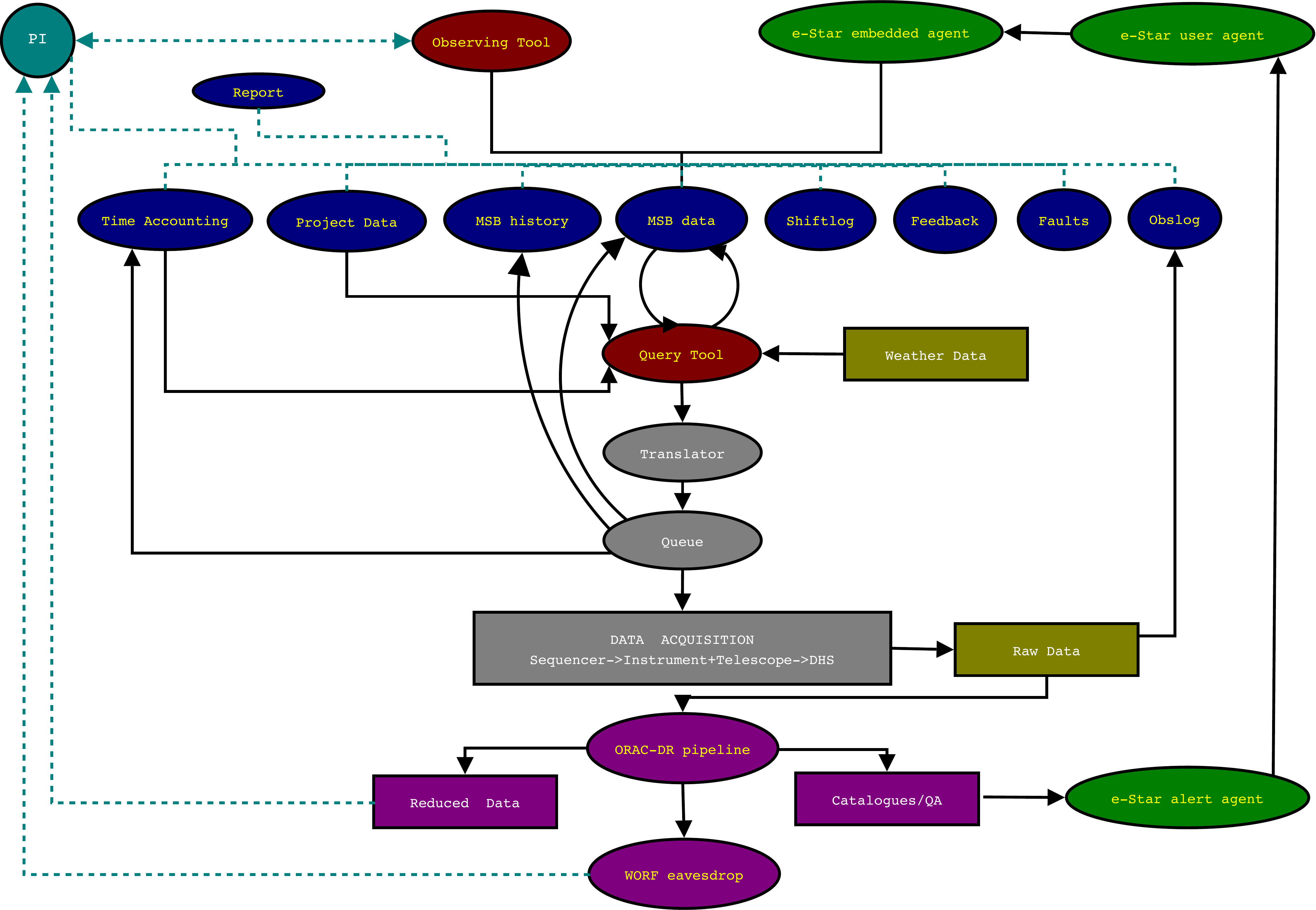}
\caption{\label{fig:omp} OMP and eSTAR system architecture.}
\end{center}
\end{figure}

The flow chart showing how the OMP interacts with eSTAR is shown in
Fig.\ \ref{fig:omp}.

\section{DATA ARCHIVING}

Given history and the different funding models for UKIRT and JCMT
these telescopes have different data archive centres. This section
describes each telescope in turn.

\subsection{JCMT}

The JCMT archives all data at the Canadian Astronomy Data Centre. We
use the standard CADC e-Transfer system initially developed for the
Gemini Science Archive\cite{2005ASPC..347..647M}. The data are
transferred to CADC within a few minutes of the files appearing on
disk. The key design issue is that metadata are sent to CADC before
the file itself is transferred and CADC reject the file if the
metadata does not match a pre-defined data dictionary. This allows
file header problems to be spotted quickly and files must be fixed by
the telescope support staff before attempting to re-ingest them.

A single JCMT observation can consist of multiple files and with
SCUBA-2 eight separate data acquisition computers write out files in
parallel every 30 seconds during an observation. These files have a
shared control task which allows certain headers to be synchronized to
make it simple to work out that they are related files. As each file
appears on disk a daemon checks the data directory by looking in a
small text file, called a ``flag'' file, to see if any new files have
been created. This flag file is only updated when the data acquisition
has finished writing a file and it is ready to be picked up by the
data reduction pipeline or archiving software. The archive daemon then
reads the header from the file and decides whether this is a new
observation or part of an existing observation.

The JCMT data archive consists of four database
tables. \texttt{COMMON} contains the metadata common to every
instrument. As well as the items from the file headers we also store
the details of the area of sky covered by the observation. There are
then instrument-specific tables which at JCMT consists of a
\texttt{SCUBA2} table and an \texttt{ACSIS} table. Theses tables can
have multiple rows per observation, known as ``sub-systems'', because
with SCUBA-2 we write data at two wavelengths simultaneously and for
the ACSIS correlator we can write out up to four separate spectra from
different positions in the 2\,GHz band of the frontend. Finally there
is a table, \texttt{FILES}, containing the names of the files
associated with each row in \texttt{COMMON} and
\texttt{SCUBA2}/\texttt{ACSIS}. The rows are linked with an
observation identifier (known as an ``obsid'') that is unique for each
observation and a subsystem identifier (``obsidss'') which is a
combination of the obsid and the subsystem number. The obsid takes
the form ``\textit{instrument}\_\textit{run-number}\_\textit{date-obs}
to enable it to be unique and human-readable. An example obsid might
be \texttt{scuba2\_00071\_20100310T101813} with a corresponding
obsidss of \texttt{scuba2\_00071\_20100310T101813\_450}.

These tables are stored in a Sybase database and automatically
replicated at CADC. There is an additional table at the JCMT that
tracks the state of a particular file. This state table has states for
``file found'', ``metadata ingested'', ``row replicated'', ``copied to
etransfer'', ``arrived at CADC'' and ``error'', making it easy to
determine how efficiently the system is transferring data and whether
we have a blockage.

For a new observation a row must be placed in each of the relevant
tables. If the new file is part of an observation that is already
present in \texttt{COMMON}, any metadata associated with the end of an
observation (such as \texttt{DATE-END} or the end elevation or
azimuth) is updated and an entry is made in the \texttt{FILES} table
and the instrument table as appropriate. Every few minutes another
daemon checks with CADC to see if the local database has been
replicated correctly to CADC. If the \texttt{FILES} table at CADC
contains a file that is awaiting transfer a soft link is made into the
e-transfer directory and the file is passed to the CADC software to be
transferred to CADC. If the file is rejected for any reason the
rejected file (or its soft link) is written to a rejection directory
indicating the reason why (reasons include: a corrupt file, an empty
file, non-standard filename, incorrect metadata, or the file is
already at CADC and was not marked for replacement).

On a daily basis a further cron job runs at the telescope designed to
free up disk space on the data acquisition machines. A file can only
be deleted if CADC report that they have successfully ingested it.

\subsection{UKIRT}

UKIRT used to have two modes of operation. When WFCAM arrived the
Cassegrain instruments (CGS4, UIST, UFTI) were used some of the year
and WFCAM was used for the rest of the year. Budget cuts and a
re-focussing on survey science meant that the Cassegrain instruments
were mothballed and UKIRT now uses WFCAM full time in order to
complete the UKIDSS survey\cite{2007MNRAS.379.1599L}. The Cassegrain
archive consisted of writing data to tapes and, when disk space became
cheaper, storing it on a disk in Hilo. Every semester data tapes were
also sent to the archive in Cambridge at the Cambridge Astronomical
Survey Unit (CASU) where they were converted to FITS format.

WFCAM generates much more data than the Cassegrain instruments and the
archive works slightly differently. The raw data format from WFCAM is
in NDF and these files are copied to Hilo and placed on LTO-3
tapes. The raw data are deleted from the summit, oldest first, when
the LTO tape has been read back and verified to be correct. On a daily
basis a cron job converts the raw data to FITS format using Rice
compression and the data are retrieved by CASU. When they have
retrieved the data a flag file is written indicating that the cleanup
jobs can remove the directory.

Storing the raw data on tape has been very useful over the years. In
general we never need to read them but there have been instances where
the conversion to FITS has been buggy and we have not realised for a
month or two. The data were no longer on disk but could be retrieved
from tape, albeit slowly, and reconverted to FITS. The main issue with
our raw data archive is the changing media. Initially files were
written to LTO-1 and from mid-2008 LTO-3. The LTO standard indicates
that an LTO-4 drive does not need to read LTO-1 tapes and given that
LTO-5 is the current standard (which will read LTO-3) it is highly
likely that the LTO-1 tapes covering 2005 to 2008 will never be read,
but of course the data have been fully reduced and published by WFAU
in Edinburgh\cite{2008MNRAS.384..637H} so there is no reason to need
to read them now.

\section{ARCHIVE PROCESSING}

\begin{figure}
\begin{center}
\includegraphics[width=0.9\textwidth]{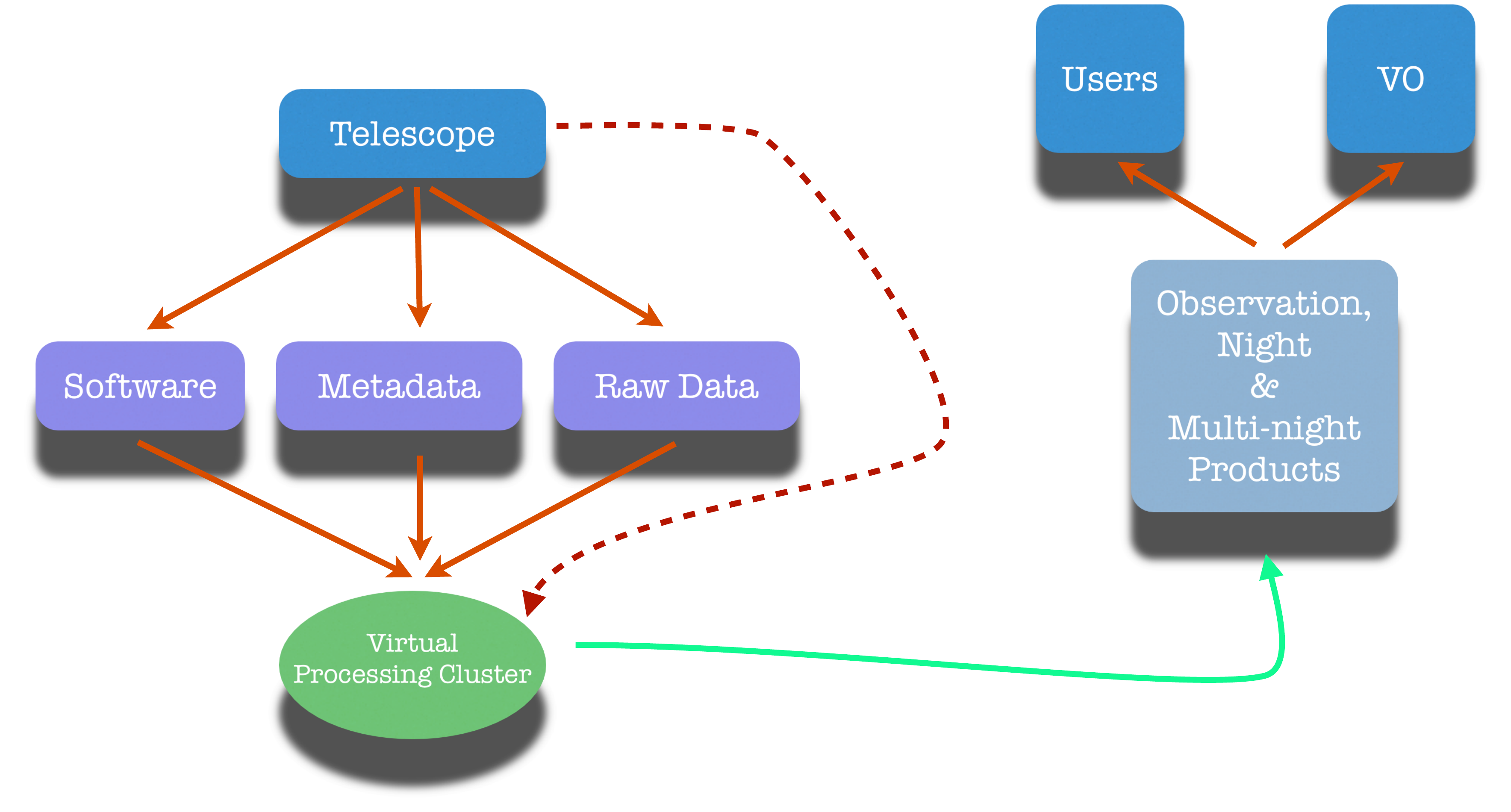}
\caption{Conceptual diagram of the JSA. The telescope makes raw data,
  metadata and archive available to the data centre. It then triggers
  processing on the data centre's processing nodes. Products are
  immediately exposed}
\label{fig:jsa-model}
\end{center}
\end{figure}

The JCMT Science Archive
(JSA)\cite{2008ASPC..394..135G,2008ASPC..394..450E,2011ASPC..442..203E} is a
collaboration between the JAC and CADC to provide calibrated reduced
products from the JCMT to the astronomy community as quickly as
possible. The JAC is interested in increasing the productivity of the
JCMT by providing its community with science-ready data products and
by enhancing the value of the JCMT Legacy Survey programme. CADC
provide their knowledge of managing large data centres and integrating
into the virtual observatory and also learn how to handle datasets
that are very different from the more common 2-D optical/IR data.

The JSA is separated into two components with well-defined
interfaces. CADC are the ``data store'' and service requests for data
files to be made available to the processing system and allow products
from the data processing to be ingested into the reduced products
database. Again metadata management is critical with CADC refusing to
accept any files that do not follow the rules. CADC also provide the
public interface to the data and handle authentication issues and
proprietary periods. The JAC are entirely responsible for data
processing and for deciding which data files should be processed at
CADC.

There are four types of products in the JSA: single observations
(``\textit{obs}''), combined observations from a single night,
combined observations from a single project and a combination of all
relevant public data for a particular field. The last product has not
yet been generated although all generated products do become public
when the proprietary period expires.

Each job submitted to the processing system is assigned a unique
``association identifier'' that links the job with the product and is
derived from the grouping algorithm used to determine that the
submitted files are related. If more observations are taken satisfying
a particular query the association identifier will not change when
those observations are included in the newly recalculated product. The
conceptual diagram of the JSA is show in Fig.\ \ref{fig:jsa-model}.

The JSA does not have periods set aside for data releases. Every time
data are reprocessed by the JSA pipeline the earlier version is
removed and replaced with the current version. Headers in the data
files indicate the versions of the software used to generate them (via
SHA1 identifiers from the \texttt{git}\footnote{http://git-scm.com/}
source code repository) along with the data processing date so it is
possible to track which version of a product has been used in a
publication. This scheme allows us to re-reduce data when there is an
update to the data pipeline. Additionally having this infrastructure
makes it very easy to test the data pipeline by doing a bulk
reprocessing on a development system.

\subsection{Provenance Tracking}

One important requirement placed on the data processing system by the
science archive is to track provenance. Every product archived in the
JSA has to have full knowledge of its parent data files and the
operations performed on that data as well as the ability to walk
through the information to determine the original raw data files that
went into the product. The NDF history mechanism\cite{sun33} only
tracks a single path through the provenance so a new system was needed
to track full provenance. Additionally, it was decided that modifying
each processing step to explicitly propagate provenance was not
acceptable so a low-level solution was required that would handle
provenance propagation automatically.

\begin{figure}
\begin{center}
\includegraphics[width=\textwidth]{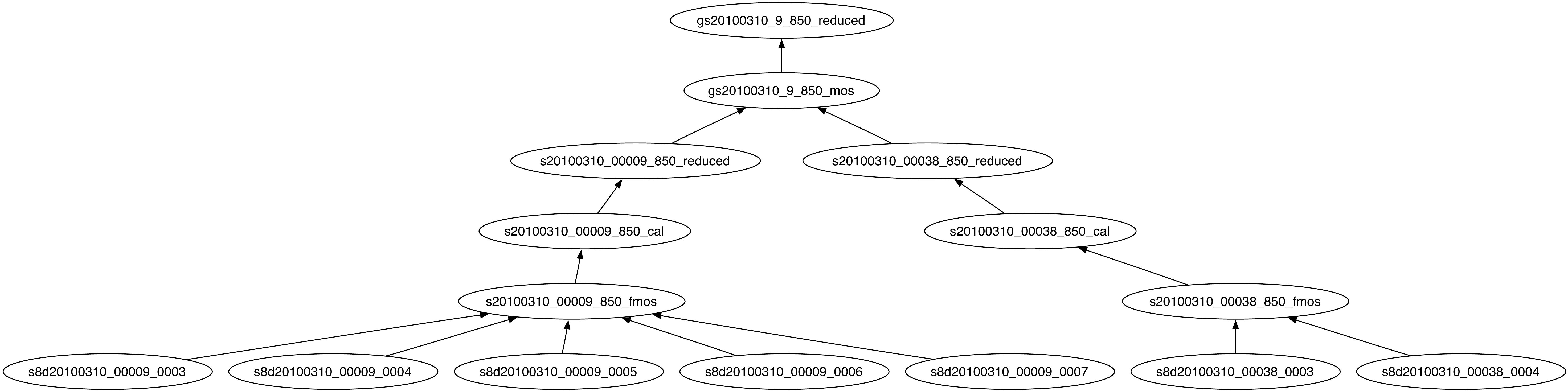}
\caption{\label{fig:prov1}
Provenance tree from two SCUBA-2 observations combined into a single
mosaicked product. There are multiple data files from each observation.}
\end{center}
\end{figure}

Most Starlink applications use the NDG library\cite{sun2} to obtain a
group of input or output files. The NDG library is a layer on top of
NDF and makes use of a system that can register callbacks with the NDF
library. When a file is opened the information is retained by the
library and when an output file is closed the provenance is propagated
to that output file unless the provenance has been explicitly
augmented by the application. This allows applications where input
data flows to output files to handle provenance transparently but more
complex applications can override with specialist knowledge. The
SMURF\cite{2011ASPC..442..281J} routines were modified to seed a root
provenance with the correct information (the ``obsidss'') and this
information then flows to the JSA products automatically. An example
provenance tree is shown in Fig.\ \ref{fig:prov1} where two SCUBA-2
observations are combined by the pipeline into a single mosaicked
image.

The pipeline provenance information includes the parent file and the
complete command details used to generate the file. The CADC
provenance infrastructure requires that the FITS header includes
details of the parent provenance and the root ancestor
``obsidss''. The parent provenance must refer to a file archived at
CADC so it is necessary to compress the provenance before ingestion by
removing temporary files from the tree.

\section{Conclusions}

The combination of the pipeline, flexible scheduling and science
archive enables the astronomer to be involved in the observing to an extent
that has been impossible since many telescopes moved to
queue-scheduled observing. The reduced data are available within a few
hours and the astronomer can adjust their science program and resubmit
it before the next observing night begins.

There is also direct feedback from the summit data quality assessment
to the archive processing. If the observer marks an observation as
questionable or bad the science archive will not combine the data into
the combined group. If the data are marked junk, hopefully a rare
occurrence, the data will not be processed at all. The science archive
pipeline itself does more detailed quality assessment of the data and
is able to mark the observation status in the OMP accordingly. This
feedback loop allows instrument scientists to get rapid feedback of
instrument issues before they become serious.

A fully automated data-driven data reduction pipeline is pivotal to
the ongoing development of the data management system at the JAC. All
new instruments must be delivered with a DR pipeline based on ORAC-DR
and must also use the standard OCS interfaces to allow simple
integration into the OMP.

As time passes UKIRT and JCMT have adopted more and more shared
infrastructure.\cite{2002SPIE.4844..321E} Starting with the release of
ORAC-DR in 1998 the JAC has continued to benefit from sharing the
software and each telescope has received updates that would not have
been possible if we had retained separate software groups.

The JSA brings together the pipeline and flexible scheduling to allow
fast feedback to astronomers giving them fully reduced observations
within a few hours of the data being taken. Many sites do not have the
computing resources required to process or even transfer the 100\,GB
of data taken each night on a reasonable timescale. This then gives
them an immediate opportunity to adjust their science programme for
subsequent nights and provides them with multi-night coadds that can
be used to jumpstart paper writing.

%%%%%%%%%%%%%%%%%%%%%%%%%%%%%%%%%%%%%%%%%%%%%%%%%%%%%%%%%%%%%
\acknowledgments     %>>>> equivalent to \section*{ACKNOWLEDGMENTS}

The James Clerk Maxwell Telescope is operated by the Joint Astronomy
Centre on behalf of the Science and Technology Facilities Council of
the United Kingdom, the Netherlands Organisation for Scientific
Research, and the National Research Council of Canada. The United
Kingdom Infrared Telescope is operated by the Joint Astronomy Centre
on behalf of the Science and Technology Facilities Council of the
United Kingdom.

%%%%%%%%%%%%%%%%%%%%%%%%%%%%%%%%%%%%%%%%%%%%%%%%%%%%%%%%%%%%%
%%%%% References %%%%%

\bibliography{tfa}   %>>>> bibliography data in report.bib
\bibliographystyle{spiebib}   %>>>> makes bibtex use spiebib.bst

\end{document}